\documentstyle[epsf,psfig,12pt]{article}
\def\Eq#1{{Eq. (\ref{#1})}}
\def\Eqs#1#2{{Eqs. (\ref{#1}) e (\ref{#2})}}
\def\fig#1{{Fig. (\ref{#1})}}
\def\gsim{\raise0.3ex\hbox{$\;>$\kern-0.75em\raise-1.1ex\hbox{$\sim\;$}}}
\def\lag{{\cal L}}
\def\ifmath#1{\relax\ifmmode #1\else $#1$\fi}
\def\half{\ifmath{{\textstyle{1 \over 2}}}}
\def\vev#1{\left\langle #1\right\rangle}
\def\vb#1{\vbox to #1 pt{}}
\def\pl#1#2#3{           {\it Phys. Lett. }{\bf #1} (19#2) #3}
\def\prl#1#2#3{          {\it Phys. Rev. Lett. }{\bf #1} (19#2) #3}
\def\np#1#2#3{           {\it Nucl. Phys. }{\bf #1} (19#2) #3}
\def\pr#1#2#3{           {\it Phys. Rev. }{\bf #1} (19#2) #3}
\def\ppnp#1#2#3{           {\it Prog. Part. Nucl. Phys. }{\bf #1} (19#2) #3}
\def\etal{\hbox{\it et al., }}
\begin{document}
\thispagestyle{empty}
\begin{titlepage}
\rightline{hep-ph/9612311}
\vskip 0.3cm
\begin{center}
{\Large \bf Limits on the $\nu_{\tau}$ mass from nucleosynthesis in the
presence of annihilation into majorons}\footnote{Work done 
in collaboration with A.D. Dolgov (Denmark), S. Pastor
and J.W.F Valle (IFIC, Val\`encia)}
\vskip 1cm
\end{center}
\normalsize
\vskip1cm
\begin{center}
{\Large \bf  J. C. Rom\~ao}
\footnote{E-mail fromao@alfa.ist.utl.pt}\\
{\it Instituto Superior T\'ecnico, Departamento de F\'{\i}sica\\
A. Rovisco Pais, 1 1096 Lisboa Codex, Portugal}
\vskip 0.5cm
Contribution to the International Workshop\\
{\it New Worlds in Astroparticle Physics}\\ 
Faro, Portugal, 8-10 September 1996.
\end{center}
\vskip 1.5cm
\begin{abstract}
We show that in the presence of sufficiently strong $\nu_{\tau}$
annihilations into majorons, the primordial nucleosynthesis constraints can
not rule out $\nu_{\tau}$ masses in the MeV range.

\end{abstract}
\end{titlepage}

\section{Introduction}

Despite great experimental efforts the tau-neutrino still remains as the
only one that can have a mass in the MeV range. The present experimental
limit on its mass is \cite{aleph}
\begin{equation}
m_{\nu_{\tau}} < 23\ MeV
\end{equation}
Further progress will have to wait for the tau-charm or B-factories. On the
other hand, many particle physics models of massive neutrinos lead to a tau
neutrino with mass in the MeV range. Moreover such a neutrino may have
interesting cosmological implications. It is therefore important to examine
in more detail the cosmological constraints.

The first comes from the critical density argument. From the contribution of
stable $\nu_{\tau}$ to the present relic density one gets $m_{\nu_{\tau}}
< 92 \Omega h^2\ eV$, where $h$ is related to the Hubble constant
through $h = H_0 / (100\ kms^{-1}\ Mpc^{-1})$. This means
that a massive $\nu_{\tau}$ with mass in the MeV range must be unstable with
lifetimes smaller than the age of the Universe. However, it has been shown
that in many particle physics models where neutrinos acquire their mass by
the spontaneous breaking of a global lepton number symmetry\cite{swieca}
this limit can be avoided due to the existence of fast $\nu_{\tau}$ decays.

The second constraint comes from primordial nucleosynthesis considerations.
In the standard model, these rule out $\nu_{\tau}$ masses in the
range\cite{KT,DR} 
\begin{equation}
0.5\ MeV < m_{\nu_{\tau}} < 35\ MeV
\label{KTDR}
\end{equation}
It is possible to weaken the constraints on $\nu_{\tau}$ of \Eq{KTDR}, by
adding new interactions beyond the standard ones. One may either consider
that the $\nu_{\tau}$'s are unstable during nucleosynthesis\cite{Steigman}
or that they possess new channels of annihilation beyond the standard ones.
It is this last possibility that we investigate in the case where the
$\nu_{\tau}$'s can annihilate to majorons (J).

\section{Evolution of $\nu_{\tau}$ number density}

\subsection{Before weak decoupling}

\begin{figure}
\centerline{\protect\hbox{\psfig{file=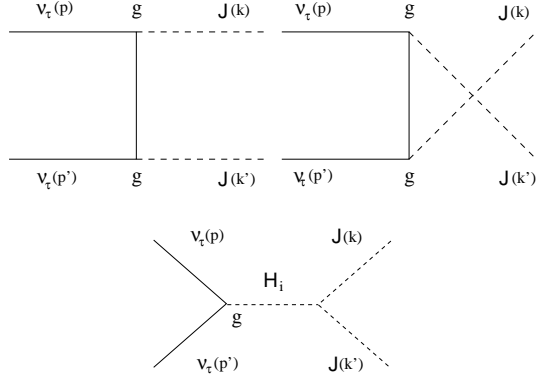,height=5cm}}}
\caption{Feynman diagrams for annihilations  of tau neutrinos
into Majorons.} 
\label{fig1}
\end{figure}

We will assume that the massive Majorana $\nu_{\tau}$'s are stable during the
nucleosynthesis epoch. They interact with leptons via the standard weak
interactions $\nu_{\tau} \overline{\nu_{\tau}} \leftrightarrow \nu_0
\overline{\nu_0}, e^+ e^-$. 
Moreover the $\nu_{\tau}$'s annihilate to majorons $\nu_{\tau}
\nu_{\tau} \to J J$ via the diagrams shown in \fig{fig1}.
The t-channel diagram is present in all Majoron models,
while the strength of the s-channel scalar exchange diagram 
is somewhat model-dependent. For the range of $g$ values 
relevant for our purposes ($g \gsim 10^{-5}$) it is quite 
reasonable to  neglect the s-channel process, especially in 
Majoron models where the breakdown of the global lepton number 
symmetry happens at the electroweak scale or higher, like 
the simplest seesaw model \cite{CMP}. We have shown\cite{bbnpaper} 
that for models
with low scale breaking of lepton number, such as given in ref. 
\cite{lowscalemodels,rpsusy} the bounds 
obtained by neglecting the s-channel process could be 
relaxed by at most a factor $\sim 2$ or so.
Therefore in what follows we will consider that
the $\nu_{\tau}$'s annihilate to
majorons $\nu_{\tau} \nu_{\tau} \to J J$ via the diagonal coupling
\begin{equation}
\lag_{\nu_{\tau} \nu_{\tau} J}
=i\ \half\ g\ J\ \overline{\nu_{\tau}}\ 
\gamma_5\ \nu_{\tau} 
\label{lagint}
\end{equation}

The evolution of the number densities of $\nu_{\tau}$ ($n_{\tau}$) and
majorons ($n_J$) is given by the solution of a set of Boltzmann differential
equations 
\begin{equation}
{\dot n}_{\tau} + 3 H n_{\tau} = - \sum_{i=e,\nu_0} \vev{\sigma_i v} 
(n^2_{\tau} - ( n^{eq}_{\tau})^2 ) 
- \vev{\sigma_J v } (n^2_{\tau} - ( n^{eq}_{\tau})^2 
\frac{n^2_J}{(n^{eq}_J)^2})\equiv S_{\nu_{\tau}} 
\end{equation}
\begin{equation}
{\dot n}_J + 3 H n_J = \vev{\sigma_J v } (n^2_{\tau} - ( n^{eq}_{\tau})^2 
\frac{n^2_J}{(n^{eq}_J)^2}) \equiv S_J 
\label{boltzmann}
\end{equation}
where $\vev{\sigma_i v}$ are the thermally averaged cross sections times the
$\nu_{\tau}$ relative velocity $v$. The cross sections are for
annihilation to majorons
\begin{equation}
\sigma_J(\eta)=\frac{g^4}{128 \pi}\
\frac{1-\eta}{m^2_{\nu_{\tau}}\eta}\ \left[ \ln \left( 
\frac{1+\sqrt{\eta}}{1-\sqrt{\eta}} \right) -2 \sqrt{\eta} \right]
\label{sigJ}
\end{equation}
and for annihilation into {\it massless} fermions 
\begin{equation}
\sigma_i(\eta)=\frac{32G^2_F}{3\pi}\ 
\frac{m^2_{\nu_{\tau}}\sqrt{\eta}}{1-\eta}\ (b^2_{Li}+b^2_{Ri})
\label{sigff}
\end{equation}
where $b^2_L+b^2_R=1/2$ for $i=\nu_0$ and $b^2_L+b^2_R=2 \left[ (-1/2+\sin^2
\theta_W )^2 +(\sin^2 \theta_W)^2 \right] \simeq 0.25$ for $i=e$. In
\Eq{boltzmann} we assumed Boltzmann statistics and $n_i=n_i^{eq}$ for
$i=e,\nu_0$. 

Now let us describe our calculations. First we normalized the number
densities to the number density of massless neutrinos, $n_0\simeq 0.181
T^3$, introducing the quantities $r_{\alpha}\equiv n_{\alpha}/n_0$, where
$\alpha=\nu_{\tau},J$, and the corresponding equilibrium functions
$r^{eq}_{\alpha}$.  Then from \Eq{boltzmann} we get
\begin{equation}
\frac{d r_{\alpha}}{d T}=\left(\frac{S_{\alpha}}{n_0}-3 H r_{\alpha}
\right)\ \frac{1}{\dot T} - \frac{3}{T}\ r_{\alpha}
\end{equation}
On the other hand, the time derivative of the temperature is obtained from
Einstein's equation
\begin{equation}
\dot \rho=-3H(\rho +P)
\label{einstein}
\end{equation}
where $\rho$ is the total energy density and $P$ is the pressure.

\subsection{Past weak decoupling}

Once the $\nu_{\tau}$'s decouple from the standard weak interactions, they
remain in contact only with the majorons. Then one has two different
plasmas, one formed by $\nu_{\tau}$'s and $J$'s and the other by the rest of
particles, each with its own temperature, denoted by $T$ and $T_{\gamma}$,
respectively. We assume that the photon temperature evolves in the usual
way, $\dot y = H y$. The evolution equations of the $\nu_{\tau}$ and $J$
number densities are now simplified versions of \Eq{boltzmann}, because we
$S_{\nu_{\tau}}=-S_J$,
\begin{eqnarray}
{\dot n}_{\tau} + 3 H n_{\tau} &=& - S_J \cr
\vb{20}
{\dot n}_J + 3 H n_J &=& S_J
\label{boltzmann2}
\end{eqnarray}
To solve these equations one must find a relation between $T$ and
$T_{\gamma}$. This is obtained using Einstein's equation for the
$\nu_{\tau}+J$ plasma. 

In order to determine the final $\nu_{\tau}$ frozen density which will be
relevant during nucleosynthesis we have to solve numerically the appropriate
set of coupled differential equations for each pair of values
($m_{\nu_{\tau}},g$). Before weak decoupling these are 
\Eqs{boltzmann}{einstein} while after decoupling one should use
\Eq{boltzmann2}. The initial conditions are, for sufficiently high 
temperatures, $r_{\nu_{\tau}}=r^{eq}_{\nu_{\tau}}$, $r_J=r^{eq}_J$ and
$T_{\nu_{\tau}}=T_J=T_{\nu_0}$.

\section{Nucleosynthesis constraints}

The value of $r_{\nu_{\tau}}(m_{\nu_{\tau}},g)$ is used to estimate the
variation of the total energy density $\rho_{tot}=\rho_R
+\rho_{\nu_{\tau}}$. In $\rho_R$ all relativistic species are taken in
account, including the majorons and two massless neutrinos, whereas
$\rho_{\nu_{\tau}}$ is the energy density of the massive $\nu_{\tau}$'s.

In order to compare with the standard model situation, we can now express
the effect of the $\nu_{\tau}$ mass and of the annihilation to majorons in
terms of an effective number of massless neutrino species, $N_{eq}$, which we
calculate for each value of $r_{\nu_{\tau}}(m_{\nu_{\tau}},g)$. To do this,
we first numerically calculate the evolution of the neutron fraction, $r_n$,
by varying the value of $N_{eq}$\cite{Dicus}. Then we incorporate
$\rho_{tot}$ in this numerical code and perform the integration for each
pair of $(m_{\nu_{\tau}},g)$ values. Comparing the $r_n$ obtained in each case
at $T_{\gamma}\simeq0.065\ MeV$ (the moment when practically all neutrons are
wound up in ${}^4He$), we can relate $(m_{\nu_{\tau}},g)$ and $N_{eq}$.

\begin{figure}
\centerline{\protect\hbox{\psfig{file=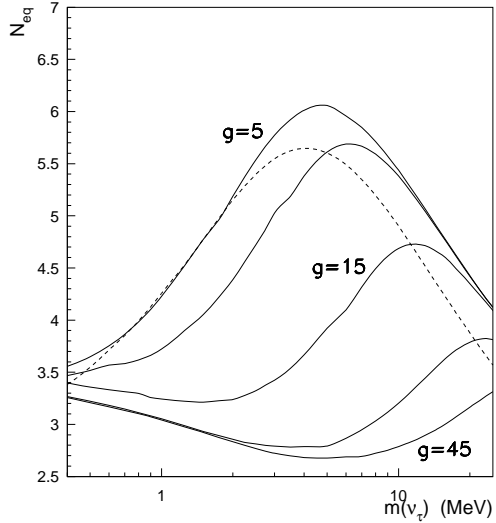,height=7cm}}}
\caption{Effective number of massless neutrinos equivalent to
the contribution of heavy $\nu_{\tau}$'s  with different values of $g$ 
 in units of $10^{-5}$. For comparison, the dashed line
corresponds to the standard model case when $g=0$.}
\label{fig2}
\end{figure}

\begin{figure}
\centerline{
\psfig{file=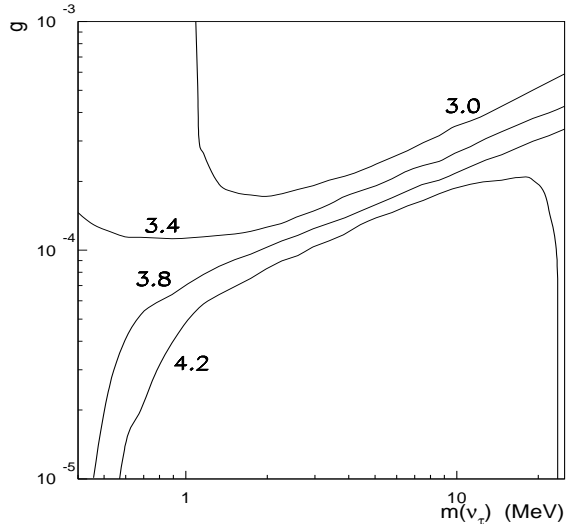,width=7.5cm,height=7cm}}
\caption{The values of $g(m_{\nu_\tau})$ above each line 
would be allowed by nucleosynthesis if one adopts the $N_{eq}^{max}
= 3, 3.4, 3.8, 4.2$ (from top to bottom). }
\label{fig3}
\end{figure}

The results are shown in \fig{fig2}. One can see that for fixed $N_{eq}$, a
wide range of tau neutrino masses is allowed for large enough coupling
constants. In fact all masses below 23 MeV are allowed provided 
the coupling constant $g$ exceeds a value a few times $10^{-4}$. For
comparison, the dashed line corresponds to the case when $g=0$ and no
majorons are present.
These results can also be expressed in the $m_{\nu_{\tau}} - g $ plane, 
as shown in \fig{fig3}. 

\section{Majoron models}

There has been a variety of majoron models proposed in the 
literature\cite{faessler}. They are attractive extensions of the standard
model where neutrinos acquire mass by virtue of the spontaneous violation of
a global lepton number symmetry. Apart from a phenomenological interest of
their own, majoron models offer the possibility of loosening the
cosmological limits on the neutrino masses, either because neutrinos decay
or annihilate to majorons.

\begin{figure}
\centerline{
\psfig{file=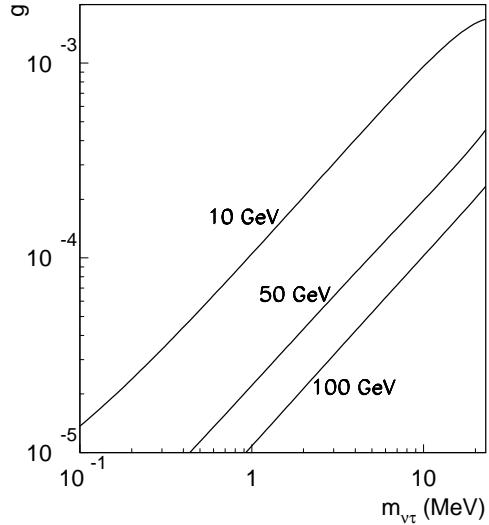,height=7cm}}
\caption{Expected values of $m_{\nu_\tau}$ and $g$ in model
of ref [9]}
\label{fig4}
\end{figure}

We have seen above that the restrictions imposed by primordial
nucleosynthesis upon a heavy tau neutrino disappear for sufficiently 
large values of the $\nu_{\tau}$ $\nu_{\tau}$ majoron coupling $g$.
Different models imply different expectations for the coupling $g$ and for
the relation between $g$ and the $\nu_{\tau}$ mass. Just to give a concrete
example let us consider the supersymmetric models with spontaneous violation
of R parity\cite{rpsusy}. In these models we get the relation between $g$
and $m_{\nu_{\tau}}$ depicted in \fig{fig4}.
The different curves correspond to different values of the parameter $v_R$
of the model. We see that we can obtain for $g$ values of the order of a few
times $10^{-4}$, as required in our nucleosynthesis analysis.

\section{Conclusions}

We have determined the restrictions imposed by  primordial nucleosynthesis
upon a heavy tau neutrino in the presence of sufficiently strong
$\nu_{\tau}$ annihilations into majorons. We have shown that for values of
the $\nu_{\tau}$ $\nu_{\tau}$ majoron coupling in excess of $10^{-4}$ all the
masses up to the experimental limit of 23 MeV can be allowed. We have
shown that in a supersymmetric model with spontaneous violation of R parity
such values can be obtained.

\bibliographystyle{ansrt}

\begin{thebibliography}{99}

\bibitem{aleph}
D. Buskulic et al., \pl{B349}{95}585.

\bibitem{swieca}
For a recent review see J.W.F. Valle, in {\it Physics Beyond the Standard
Model}, lectures given at the {\it VII Jorge Andr\'e Swieca Summer School}
(Rio de Janeiro, February 1995) and at {\it V Taller Latinoamericano de
Fenomenologia de las Interacciones Fundamentales} (Puebla, Mexico, October
1995); {\tt hep-ph/9603307}.

\bibitem{KT}
E. W. Kolb, M.S. Turner, A. Chakravorty and D. N. Schramm, \prl{67}{91}{533}.

\bibitem{DR}
A. D. Dolgov and I. ZA. Rothstein, \prl{71}{93}{476}


\bibitem{Steigman}
M. Kawasaki, G. Steigman and H.-S. Kang, \np{B402}{93}{323}, 
\np{B419}{94}{105}; S. Dodelson, G. Gyuk and M.S. Turner, \pr{D49}{94}{5068}.

\bibitem{CMP}
Y. Chikashige, R. Mohapatra and R. Peccei, \prl{45}{80}{1926}.

\bibitem{bbnpaper}
A. D. Dolgov, S. Pastor, J.C. Rom\~ao and J. W. F. Valle, 
{\tt hep-ph/9612311}.

\bibitem{lowscalemodels} 
M.C. Gonz\'alez-Garc\'{\i}a and J.W.F. Valle, \pl {B216} {89} {360};
M.C. Gonz\'alez-Garc\'{\i}a, A. Santamaria and J.W.F. Valle, 
\np{B342}{90}{108};
A. Joshipura and  J.W.F. Valle, \np{B397}{93}{105};
J.T. Peltoniemi, and J.W.F. Valle, \pl{B304}{93}{147};
Z. Berezhiani, A.Yu. Smirnov and J.W.F. Valle, \pl{B291}{92}{99};
J.C. Rom\~ao,  F. de~Campos, and  J.W.F. Valle, \pl{B292}{92}{329}.

\bibitem{rpsusy}
A. Masiero and J. W. F. Valle, \pl{B251}{90}{273}; J. C. Rom\~ao, C. A.
Santos and J. W. F. Valle, \pl{B288}{92}{311}.

\bibitem{Dicus}
D.A. Dicus \etal, \pr{D26}{82}{2694}.

\bibitem{faessler}
For recent reviews see J. W. F. Valle, {\it Gauge Theories and the Physics
of Neutrino Mass}, \ppnp{26}{21}{91-171} (ed. A. Faessler)



\end{thebibliography}

\end{document}